\documentstyle[oldlfont,12pt]{article}
\long\def\del#1\enddel{ } 
\let\3=\ss \catcode`\"=\active \let"=\"
\let\a=\alpha \let\b=\beta  \let\d=\delta \let\e=\varepsilon
  \let\th=\theta  
\let\l=\lambda    \let\p=\pi \let\r=\rho
\let\s=\sigma \let\t=\tau  
  \let\Ph=\phi
    
   \let\D=\Delta

\def\0{\over }    \def\1{\vec }  \def\2{{1\over2}} \def\3{{\ss}}
\def\4{{1\over4}} \def\5{\bar }  \def\6{\partial } \def\7#1{{#1}\llap{/}}
\def\8#1{{\textstyle{#1}}}        \def\9#1{{\bf {#1}}}

\def\<{\langle } \def\>{\rangle }  
 
\def \({\left( } \def \){\right) }

  \let\ex=\times   
      \let\and=\wedge
     
\def\|#1{{}_{\big|_{#1}}}

\def\mao#1{\mathop{\rm {#1}}\nolimits}   
  \def\mod{\mao{mod}}

\def\pmbf#1{\setbox0=\hbox{${#1}$}   \kern-.025em\copy0\kern-\wd0
      \kern.05em\copy0\kern-\wd0     \kern-.025em\raise.0433em\box0 }

 \def\co{{\cal O}} 
\def\inbar{\vrule height1.5ex width.4pt depth0pt}
\def\IN{\relax{\rm I\kern-.18em N}}    \def\ZZ{\relax{\sf Z\kern-.4em Z}}
\def\IP{\relax{\rm I\kern-.18em P}}
\def\IQ{\relax\,\hbox{$\inbar\kern-.3em{\rm Q}$}}
\def\IR{\relax{\rm I\kern-.18em R}}
\def\IC{\relax\,\hbox{$\inbar\kern-.3em{\rm C}$}}

\def\beq{\begin{equation}} \def\eeq{\end{equation}}
\def\eql#1{\label{#1}\eeq} \def\bea{\begin{eqnarray}}
\def\eea{\end{eqnarray}}  

\def\plb#1 #2 {Phys. Lett. {\bf B#1} #2 }
\def\phr#1 #2 {Phys. Rep. {\bf  #1} #2 } 
\def\npb#1 #2 {Nucl. Phys. {\bf B#1} #2 }
\def\aph#1 #2 {Ann. Phys. {\bf #1} #2 }  
\def\jmp#1 #2 {J. Math. Phys. {\bf #1} #2 }
\def\prd#1 #2 {Phys. Rev. {\bf D#1} #2 }
\def\prl#1 #2 {Phys. Rev. Lett. {\bf #1} #2 }
\def\rmp#1 #2 {Rev. Mod. Phys.  {\bf #1} #2 }
\def\zpc#1 #2 {Z. Phys. {\bf #1C} #2 }
\def\cmp#1 #2 {Comm. Math. Phys. {\bf #1} #2 }
\def\mpl#1 #2 {Mod. Phys. Lett. {\bf A#1} #2 }
\def\ijmp#1 #2 {Int. J. Mod. Phys. {\bf A#1} #2 }

     \def\C#1){\IC_{#1)}}

\newcommand{\fll}     {\mbox{$\phi^{l,q,s}_{\bar l,\bar q,\bar s}$}}
\newcommand{\forien}  {\mbox{for all $i=1,2,...\,,N$}}
\newcommand{\ii}      {\mbox{i}}
\newcommand{\Lg}      {Lan\-dau--Ginz\-burg} \let\LG=\Lg
\newcommand{\lgo}     {Lan\-dau--Ginz\-burg orbifold}
\newcommand{\ns}      {\mbox{$n^{}_{27}$}}
\newcommand{\nsb}     {\mbox{$n^{}_{\overline{27}}$}}
\newcommand{\otime}   {\!\otimes\!}
\newcommand{\pp}      {Poin\-car\'e polynomial}
\newcommand{\zz}[1]   {\mbox{$\ZZ_{#1}$}}

\begin{document} 
\begin{centering}\mbox{}\\[19mm]  $\!\!${\Large\bf
 On the \Lg\ description of $(A_1^{(1)})^{\oplus N}_{}\!$ invariants}$\!\!$%
\\[11mm] {\large \ J\"urgen Fuchs $^*$}\\[4 mm] NIKHEF-H\\ Kruislaan 409,
\ NL\,--\,1098\ \ SJ Amsterdam\\[4 mm]  and\\[4 mm] {\large Maximilian
Kreuzer} \\[5 mm] Theory Division\\CERN\\
     CH\,--\,1211\ \ Gen\`eve 23       \\[1 cm] {~}
\end{centering}
\begin{quote}{\bf Abstract}

We search for a \Lg\ interpretation of non-diagonal modular invariants
of tensor products of minimal $n=2$ superconformal
models, looking in particular at automorphism invariants and at some
exceptional cases. For the former we find a simple
description as \lgo s, which reproduce the correct chiral rings as well
as the spectra of various Gepner--type models and orbifolds thereof.
On the other hand, we are able to prove for one of the exceptional cases
that this conformal field theory cannot be described by an orbifold of
a \Lg\ model with respect to a manifest linear symmetry of its potential.
 \\[23mm] {~\vrule width0pt} \end{quote}
  {\vspace{11 mm}\noindent{\sf CERN-TH.6669/92\\October 1992}\\[7 mm]
-----------------\\[2 mm] $^*$\ \,Heisenberg Fellow

   \vspace{-24.9 cm}\rightline{\sf CERN-TH.6669/92}
   \newpage \mbox{} \setcounter{page}{0} \textheight=235truemm
\newpage

    \section{Introduction}

Compactifications of the heterotic string to four space-time dimensions
have been analysed not only because of their 
prospects for model building (see e.g.\ \nocite{rr,more,sch}%
{\hbox{\makeatletter[\@nameuse{b@rr}--\@nameuse{b@sch}]\makeatother}}),
but also from a less phenomenological point of
view, as they provide a connection between otherwise rather distinct
areas of research such as Calabi--Yau manifolds \cite{cy}, \Lg\ theories
\cite{mvw,lvw}, and exactly solvable $n=2$ superconformal two-dimensional
field theories \cite{gep}. An issue that is of relevance to both the
phenomenological and the more mathematical aspects is to know to
which extent the sets of compactified string models obtained via the
various approaches overlap or even are contained in each other.

More specifically, one may ask whether all models which arise from
`compactification' on tensor products of minimal $n=2$ superconformal
field theories -- the so-called Gepner--type \cite{gep0} models -- possess
a \Lg\ interpretation.  It is this question that we are going to address
in the present paper.  A comparison between the space of \Lg\ theories
and the one of Gepner--type models
is desirable for the following reasons. On the one hand,
in the \Lg\ framework it is a lot easier than with conformal field theory
methods to compute spectra of massless (gauge non-singlet) string modes,
and hence to search for interesting\,\footnote{~Here the
qualification `interesting' may be taken in the sense that the
model possesses a phenomenologically promising low-energy limit
\cite{sch}, but it may concern different aspects as well.  For example,
in the context of `mirror symmetry' \cite{ms,bh} one may look for the
mirror partner of a particular model.}
 models by scanning large classes of string compactifications.
Thus if {\em all\/} Gepner--type models possessed a \Lg\
interpretation, it would not be necessary to scan them
separately. On the other hand, after having found a \Lg\ theory of
interest, one could employ the conformal field theory
machinery to obtain more detailed information about the model, e.g.\
compute Yukawa couplings with the help of operator product expansions.
In addition, if the \Lg\ description of some Gepner--type model is in
terms of an orbifold of a Fermat--type \Lg\ potential, it is not even
necessary to treat this model independently in the conformal field theory
framework, but rather one may describe it as the corresponding orbifold
of a Gepner--type model which involves diagonal modular invariants only.

More generally, it is always convenient to have two different
descriptions of a particular model. Looking at the model from different
points of view, one can gain new insight in its
structure, and possibly even in the nature of the methods on which
the descriptions are based.

So far the issue raised above has been analysed
only for models which employ the standard
$A$-$D$-$E$--type modular invariants of $A_1^{(1)}$. The result was that
all these models possess a simple \Lg\ description: for $A$--type
(diagonal) invariants as well as for the $E_6$ and $E_8$ invariants one
is dealing with Fermat--type potentials \cite{mvw,lvw}, and for $D$--type
invariants with $\ZZ_2$ orbifolds thereof \cite{ls}, while
the $E_7$ invariant corresponds to a non-Fermat type potential.
In this paper we extend the analysis to modular invariants of
$(A_1^{(1)})^{\oplus N}$ with $N\geq2$, which cannot be written as
products of $A$-$D$-$E$--type invariants. Such invariants have been
described in \nocite{sy,s,f} 
{\hbox{\makeatletter[\@nameuse{b@sy}--\@nameuse{b@f}]\makeatother}}.

As it turns out, there exist large classes of such invariants for which
we can identify a corresponding \lgo. However, we are also able to
find theories which
definitely cannot be described by \Lg\ theories or their orbifolds
with respect to manifest linear symmetries of their potentials.
Our paper is organized as follows. In section 2 we gather the
background information about $n=2$ superconformal models and about
\Lg\ theories which is needed below. Afterwards we present,
in section 3, invariants that cannot be expressed in the \Lg\
framework. Section 4 deals with invariants that we are able to
identify with \lgo s. At the end of that section we present a few
spectra of compactified string theories which employ these invariants.
Finally, some open questions are mentioned in section 5.

\section{Minimal $n=2$ superconformal models and \Lg\ theories}

To compare Gepner--type and \Lg--type string compactifications, one may
proceed via the identification of massless string modes, or more
specifically, of the numbers \ns\ and \nsb\ of massless modes
transforming in the
two inequivalent 27-dimensional representations of the $E_6$ part of
the gauge group. In Calabi--Yau language, \ns\ and \nsb\ correspond to
the Hodge numbers $h_{1,2}$ and $h_{1,1}$, respectively, so that
$\chi=2(\nsb-\ns)$ is the Euler number of the manifold.

However, the comparison actually need not be made for the full
compactified string models; rather, it already should be performed for the
individual $n=2$ superconformal minimal models which are the building
blocks of the Gepner--type compactifications.  The objects of interest are
then the members of the 
(chiral,\,chiral) ring \cite{lvw}.\,\footnote{~We could, instead,
 work just as well with the Ramond ground states.}
In conformal field theory terms, these are the primary conformal fields
that are annihilated by the modes $G^+_{-1/2}$ and $\bar G^+_{-1/2}$ of
the supercurrents, the chiral primary fields. They may also
be characterized by the fact that their charges $Q$ and $\bar Q$ with
respect to the $U(1)$ currents $J$ and $\bar J$ contained in the
holomorphic and antiholomorphic $n=2$ algebras
are related to their conformal dimensions $\Delta$ and $\bar\Delta$ by
  \beq Q=2\Delta, \quad \bar Q=2\bar\Delta.  \eeq
Below we briefly describe how chiral primary fields are realized in
minimal models and in \Lg\ theories.

Consider first the $n=2$ superconformal minimal models. They
are conveniently described in terms of the coset construction
$(A_1^{(1)})_{k}^{}\oplus (u_1)_2^{} /(u_1)_{k+2}^{}$. Correspondingly
any $N$-fold tensor product of minimal models can be characterized by
a set $(k_1,k_2,...\,,k_N)$ of positive integers (the levels of the
corresponding affine algebras $A_1^{(1)}$) together with some
modular--invariant combination of $(A_1^{(1)})^{\oplus N}$\, characters, and
the conformal central charge is given by $c=\sum_{i=1}^N3k_i/(k_i+2)$.
The primary fields of each minimal model are of the form \cite{gep}
\fll, where the quantum numbers $l$, $q$, and $s$ are integers referring
to the highest weights of the holomorphic symmetry algebras
$(A_1^{(1)})_{k}^{}$,
$(u_1)_{k+2}^{}$, and $(u_1)_2^{}$, respectively, and analogously
for the antiholomorphic part. In particular, the label $l$ is constrained
by $0\leq l\leq k$; also, fields \fll\ that are
related by certain recursion relations have to be identified. The
conformal dimensions and $U(1)$ charges are simple expressions in
these quantum numbers, and hence the
requirement of chirality is easily imposed, leading in the
Neveu--Schwarz sector to the condition
  \beq  q=l, \ s=0 \qquad \mbox{or} \qquad q=-l-2, \ s=-2  \label1  \eeq
on the holomorphic quantum numbers, and to
an analogous restriction on the
antiholomorphic ones. Because of the field identifications,
for one set of the quantum numbers, say the holomorphic ones,
one can always restrict
to the first of these solutions. If the modular invariant chosen for
the affine $(A_1^{(1)})_{k}^{}$ algebra is the diagonal one, then one
obviously needs $l=\bar l$, and hence the chiral primary fields must obey
  \beq  q=\bar q=l=\bar l, \quad s=\bar s=0.   \label2 \eeq
Moreover, as long as the $u_1$ invariants are the diagonal ones, which
implies that $q=\bar q\ (\mbox{mod}\,2k+4)$ and $s=\bar s\
(\mbox{mod}\,4)$, the condition (\ref 1) still implies (\ref 2), no
matter which affine modular invariant is chosen; analogously, one
still has $q_i=\bar q_i=l_i=\bar l_i, \ s_i=\bar s_i=0$ \forien, even
if one is dealing with a $(A_1^{(1)})^{\oplus N}$\, invariant of the
non-direct
product type, which is the type of invariants that we consider in this
paper.

Now we come to the description of \Lg\ theories \cite{lvw}.
Such a theory is defined by the (super)potential $W$
of a supersymmetric two-dimensional lagrangian field theory describing
scalar superfields $\Ph_i$ ($i=1,2,...\,,I$); the potential is believed
not to be renormalized owing to $n=2$ superconformal invariance.  A large
class of conformal field theories can be described by such \Lg\ models
and their orbifolds.

If the potential $W(\Ph_i)$ is quasi-homogeneous, i.e.\ satisfies
\beq W(\l^{n_i}\Ph_i)=\l^d W(\Ph_i) \eeq
for some integers $n_i$ and $d$, then the scaling dimensions and $U(1)$
charges of the superfields $\Ph_i$ are given by $\D_i=Q_i/2$ and
$Q_i=n_i/d$, and the conformal central charge is $c=3\sum_{i=1}^I(1-2Q
_i)$. The equations of motion identify the partial derivatives $\6_iW
\equiv\6 W/\6 \Ph_i$ of the potential as descendant fields.
As a consequence, the chiral ring is isomorphic with the local algebra
of the potential, i.e.\ with the factor ring of the polynomial ring in
the fields $\Ph_i$ with respect to the ideal generated by the gradients
$\6_iW$.

In order that this ring be finite, $W$ needs to have an isolated critical
point \cite{agv}
at $\Ph=0$.  This implies, in particular, that for any field
$\Ph_i$, $i=1,...\,,I$, the potential must contain a monomial of the form
$\Ph_i^a$ or of the form $\Ph_i^a\Ph_j$. In the latter
case we say that $\Ph_i$ {\em points at} $\Ph_j$, and we
refer to the sum of these  $I$ monomials as to a {\em skeleton}
of the isolated singularity. We call the set of all non-degenerate
polynomials that are quasi-homogeneous with charges $Q_i=n_i/d$ a
{\em configuration\/} $\hbox{\,}\IC_{(n_1,\ldots,n_I)}[d]$.
 Each skeleton already determines a unique configuration and also has
only a finite group of phase symmetries.  Additional monomials in $W$,
which are required for non-degeneracy if more than one fields point at a
particular further field~\cite{ksc}, can only reduce this symmetry.  On
the other hand, a given configuration may accommodate several different
skeletons.

We will try to identify a conformal field theory with a particular orbifold
of a \LG\ model by comparing the chiral rings.  The charge degeneracies
in this ring are conveniently summarized by the so-called \pp\ $P(t,\bar
t)$, which is the polynomial in $t^{1/d}$ and $\bar t^{\, 1/d}$ defined as
the sum of $t^{ Q}\bar t^{ \bar Q}$ carried out over all chiral primary
fields, $P(t,\bar t)={\rm tr}_{({\rm c,c})} t^{ J_0}\bar t^{ \bar J_0}$.
We will see that this information is already sufficient,
in all the cases we consider,
to identify the model or disprove the existence of a \LG\
representation (under our general assumption of considering only manifest
symmetries).  If a particular orbifold has the correct \pp, we may
further check the identification by considering symmetries, which provide
selection rules for the operator product expansions.  We will do this by
calculating the massless modes of Gepner--type models that employ the
modular invariant under investigation, and of orbifolds thereof.  In all
cases where the \pp\ works out correctly we also find that the results
for the numbers \ns\ and \nsb\
of generations and antigenerations calculated in both
frameworks agree.\,\footnote{~Special care, however, is necessary to
correctly identify the action of symmetry groups in the case of orbifolds
(see section 4.2 below).} This is rather non-trivial, as the projection
onto integer charges (and with respect to further symmetries) leads to
additional twisted sectors, whose contribution to the spectrum does not
correspond to any of
the chiral states of the original conformal field theory.

For untwisted theories, the \pp\ is given by $P(t,\bar t)\equiv
P(t\bar t)$ with
  \beq P(t)=\prod_{i=1}^I\frac{1-t^{1-Q_i}}{1-t^{Q_i}} . \eeq
For orbifolds, however, this formula is of limited use, as we need to
project onto invariant states. In particular, the unique  chiral primary
field with left--right charges ($c/3,c/3$) transforms with the inverse
determinant squared of the matrix describing the transformation.  If this
state should survive, as is the case in all examples considered in this
paper, this imposes the restriction $\det g=\pm1$ on the relevant
symmetry groups.  Abelian symmetry groups can further be assumed to be
diagonalized, i.e.\ to act as phase symmetries.  They are a direct
product of cyclic groups of order $\co$, which we denote by
  \beq \ZZ_\co^{}(p_1,p_2,...\,,p_I), \eql{co}
where the integers $p_i$ in parentheses indicate that the $i^{th}$ field
transforms with a phase $\exp(2\p\ii\,p_i/\co)\!\!$ under the generator
of the group.

In addition to the projected untwisted states the spectrum of the
orbifold contains twisted sectors.  Their respective (Ramond ground
states and) chiral rings only get contributions from the untwisted
fields.  The left--right charges of the twisted (Neveu--Schwarz) vacua
$|h\>$, \beq \sum_{\th_i^h>0}(\8\2-Q_i\pm(\th_i^h-\8\2)) \eql T with
$\th_i^h$ defined by $hX_i=\exp(2\p\ii\th_i^h)X_i$ and $0\le\th_i^h<1$,
and their transformation under a group element $g$,
  \beq g|h\>=(-1)^{K_gK_h}\e(g,h)(\det g|_h^{})(\det g)^{-1}|h\>, \eql H
have been determined by Intriligator and Vafa \cite{v,iv}. In (\ref H)
$\det g|_h^{}$ is the determinant for the action
of $g$ on those fields that are invariant under
$h$, and $\e(g,h)$ are discrete torsions, satisfying 
 certain consistency conditions \cite{dt,iv}. The integer
$K_g$ fixes the sign of the action of $g$ in the Ramond sector; for our
purposes we can set $(-1)^{K_g}=\det g$ \cite{iv}.

The first step in finding a \Lg\ representation with conformal central
charge $c$ which possesses some prescribed \pp\ is to enumerate all
non-degenerate configurations with the correct central charge, i.e.\ all
sets $\,\IC_{(n_1,\ldots,n_I)}[d]$ of charges $Q_i=n_i/d$
such that there is a non-degenerate polynomial in $I$ variables which is
quasi-homogeneous with respect to these charges and for which
$c/3=\sum_i(1-2Q_i)$.  As all non-degenerate configurations with
$c=9$ have been enumerated \cite{ksn,kls}, this procedure is
straightforward once we know a single non-degenerate polynomial with
$c'=9-c$, i.e.\ with $\sum_j(1-2Q'_j)=3-c/3$.  Having thus obtained a
list of candidate configurations, we can check for each candidate whether
the \Lg\ theory or any of its orbifolds gives rise to the correct \pp;
once a candidate with the right \pp\ is found, we can further check
for the chiral ring and for string spectra.  If a candidate passes all
these tests, we consider it as the \Lg\ equivalent of the conformal field
theory.  In fact, in all cases where we succeed in finding a \LG\
description, it is in terms of an orbifold of a Fermat--type potential.
As a consequence it would be possible, although very tedious, to complete
the check of this identification by calculating the full operator product
algebra of the chiral primaries.

Let us stress that we only consider orbifolds with respect to manifest
linear symmetries of the potential.
Of course, there could be additional symmetries of the conformal field
theory, such as the non-linear
transformation that permutes a $\ZZ_2$-orbifold of $X^{2a}+U^2$ and a
$D$ invariant $Y^a+YV^2$, or even the continuous symmetries of the torus
obtained, for example, from the potential $X^3+Y^3+Z^3$ by modding the
diagonal $\ZZ_3$.
Unfortunately, one does not have a handle on such symmetries within
the usual computational framework of \LG\ models.

    \section{Invariants without \Lg\ interpretation}

For generic values of $N$ and of the levels $k_i$ there is a large
variety of $(A_1^{(1)})^{\oplus N}$\, invariants $Z_{}^{(k_1,...,k_N)}$
which are not of the direct product form $\prod_{j=1}^NZ_{}^{(k_j)}$
\cite{sy0,s}.  Among these, there are many infinite series of invariants
analogous to the $A$ and $D$ series of $A_1^{(1)}$ invariants, as well as
a small number of exceptional invariants (similar to the $E$--type
invariants of $A_1^{(1)}$) that occur at isolated values of the levels
$k_i$.  As it turns out, not 
all of these exceptional modular
invariants of products of minimal models can be described by orbifolds of
\LG\ models, at least not by orbifolds with respect to manifest linear
symmetries of the potential.

We start our discussion of $(A_1^{(1)})^{\oplus N}$\, invariants
of non-direct product form with two such
exceptional invariants.
Several of them 
can be obtained via conformal embeddings. Among these, there are the
well-known $E_6$- and $E_8$--type invariants of $A_1^{(1)}$, but also
a few others such as \cite{sy} the following invariant
of $A_1^{(1)}\oplus A_1^{(1)}$ for $k_1=3,\ k_2=8$:
\beq  
Z^{(3,8)} = | 0,\!0 \oplus 2,\!4 \oplus 0,\!8 |^2
       \oplus | 0,\!4 \oplus 2,\!2 \oplus 2,\!6 |^2 
       \oplus | 1,\!2 \oplus 1,\!6 \oplus 3,\!4 |^2
  \oplus | 1,\!4 \oplus 3,\!0 \oplus 3,\!8 |^2. 
\label y \eeq
There also exist exceptional modular invariants which cannot be
obtained from conformal embeddings.
Some invariants of this type have been found in \cite{f}; among them
there is the following $A_1^{(1)}\oplus A_1^{(1)}$ invariant
at levels $k_1=k_2=8$:
  \beq  \begin{array}{lll}
  Z_{}^{(8,8)} &=& | 0,\!0 \oplus 0,\!8 \oplus 8,\!0 \oplus 8,\!8 |^2_{}
  \oplus | 2,\!2 \oplus 2,\!6 \oplus 6,\!2 \oplus 6,\!6 |^2_{} \\[1 mm]&&
  \oplus \left[ ( 0,\!2 \oplus 0,\!6 \oplus 8,\!2 \oplus 8,\!6 \oplus
  2,\!0 \oplus 2,\!8 \oplus 6,\!0 \oplus 6,\!8 ) \otime(4,\!4)^*_{}
  \oplus \mbox{c.c.} \right]
  \\[1 mm]&& \oplus\,|2,\!4 \oplus 6,\!4 \oplus 4,\!2 \oplus 4,\!6 |^2_{}
  \oplus | 0,\!4 \oplus 8,\!4 \oplus 4,\!0 \oplus 4,\!8 |^2_{}
  \oplus 2\,| 4,\!4 |^2_{} .   \end{array} \label 8 \eeq

We will now prove that the exceptional modular invariant
(\ref y) at levels $(k_1,k_2)=(3,8)$ cannot be described by a \lgo.
This model has central charge $c/3=7/5$ and \pp\
  \beq  P(t)=1+2t^{2/5}+3t^{3/5}+3t^{4/5}+2t+t^{7/5} .  \label 7 \eeq
As discussed in the previous section, to enumerate all
non-degenerate configurations with $\sum_i(1-2Q_i)=c/3=7/5$,
we only need to make use of
a single non-degenerate polynomial with $\sum_j(1-2Q'_j)=3-7/5=8/5$,
such as $X^{10}+Y^{10}$, which corresponds to the
configuration $\IC_{(1,1)}[10]$. Searching the list \cite{ksn} of
$c/3=3$\, configurations for entries
containing $\,\IC_{(1,1)}[10]$, and eliminating those 
for which the part with $c/3=7/5$
is degenerate, we find the following 10 candidates:
$\hbox{\,}\IC_{(1,2)}[10]$, $\IC_{(1,5)}[20]$, $\IC_{(2,7)}[30]$,
$\IC_{(5,7)}[40]$,
$\IC_{(1,1,2)}[5]$, $\IC_{(5,5,2)}[15]$, $\IC_{(3,4,5)}[15]$,
$\hbox{\,}\IC_{(1,4,7)}[15]$, $\,\IC_{(2,5,9)}[20]$, and $\IC_{(4,7,9)}[25]$.
Obviously, without further orbifoldizing
none of these configurations possesses (\ref 7)
as its \pp.

Thus as a next step we have to
check whether any of the orbifolds of a model belonging to the above
configurations can reproduce the correct \pp.  Generically,
configurations involving several fields of equal weight may be difficult
to handle, owing to the presence of nonabelian symmetries.  Fortunately,
although among the configurations just listed there are two cases with two
fields of equal weight, none of them can accommodate a non-degenerate
polynomial with a nonabelian symmetry group.  To see this, note that any
such quasi-homogeneous polynomial would have to be of the form
$\sum P_i(X,Y) Z^i$, with a non-vanishing linear term
$P_{(d-n_1)/n_3}=\a X+\b Y$.
By a change of variables we can set $\b=0$. Then any linear symmetry
respecting quasi-homogeneity has to be diagonal and the symmetry group
must thus be
abelian. We can therefore restrict our considerations to phase symmetries
in all 10 cases.

A necessary ingredient for obtaining the \pp\ (\ref 7)
is to use only twists by symmetries with
determinant $\pm1$, as the unique chiral primary field of highest charge
$c/3=7/5$ transforms with the inverse determinant of the twist squared.
This very restriction, on the other hand, implies in most cases that
the untwisted sector contains invariant states with undesirable charges.

For the configuration $\IC_{(1,2)}[10]$, for example, any non-degenerate
polynomial has to contain $P_1=X^{10}+Y^5$ or $P_2=X^8Y+Y^5$ (the
coefficients have been rescaled to unity).  These two polynomials
represent the points of maximal symmetry in the moduli space of the
configuration.  The symmetries we have to consider are thus generated by
$\ZZ_2(1,0)$ and $\ZZ_5(1,4)$. In the
first case, with polynomial $P_1$, we can disregard the $\ZZ_2$, as it
would only bring us to the $D$ invariant, i.e.\ to the configuration
$\IC_{(1,1,2)}[5]$ to be considered below. Then the projection onto states
invariant under the $\ZZ_5$ group keeps the field
$XY$, which has charge $3/10$ and hence should not belong to the chiral
ring.  In the second case, on the other hand, we only have the $\ZZ_2$
symmetry, which leaves the field $Y$ with charge $1/5$ invariant.  In the
same way one can check that any orbifold (with respect to a symmetry
satisfying $\det=\pm1$) of the configurations
\,$\IC_{(1,5)}[20]$, $\IC_{(2,7)}[30]$, $\IC_{(3,4,5)}[15]$,
$\hbox{\,}\IC_{(1,4,7)}[15]$, and $\IC_{(2,5,9)}[20]$ has an invariant chiral
field of charge $1/5$, whereas for $\IC_{(5,7)}[40]$ there is an
invariant field with charge $3/10$, namely $(XY)^2$.

To exclude the configuration $\IC_{(4,7,9)}[25]$, we need to go one
step further.  In this case the non-degenerate polynomial with maximal
symmetry is $X^4Z+Y^3X+Z^2Y$ and has the symmetry $\ZZ_5(2,1,2)$ with
determinant 1.  From
the untwisted sector we thus get $P_u(t)=1+t^{3/5}+t^{4/5}+t^{7/5}$,
which is not yet in contradiction with (\ref 7).  The twisted sectors,
however, contribute four states with asymmetric charges $(Q,\bar
Q)=(1/5,6/5)$ or $(6/5,1/5)$, which again are not present in the chiral
ring of the modular invariant we want to describe.

Finally, we are left with the two configurations $\IC_{(1,1,2)}[5]$ and
$\IC_{(5,5,2)}[15]$, which are a little more tedious, as there is a larger
number of polynomials with maximal symmetry.  In the first configuration
$X_1$ and $X_2$ can point at any variable, whereas in the second
configuration $X_1$ and $X_2$ can point at each other.  As a consequence,
we get 9 and 4 different skeletons, respectively. (If two fields point at
the same additional field we need, in fact, additional monomials
for non-degeneracy \cite{ksc}, which further restricts the symmetry.) All
resulting orbifolds, however, can be excluded as candidates for
describing the modular invariant (\ref y) with the same arguments as
above.

The modular invariant $Z^{(8,8)}$ described in (\ref8) above can be
analysed in a similar manner.
The \pp\ reads
  \beq P^{(8,8)}(t)=1+3t^{2/5}+2t^{3/5}+6t^{4/5}+2t+3t^{6/5}+t^{8/5}  .
  \eql{P8}
There are now 20 non-degenerate configurations that have the correct central
charge. Ten of
these correspond to $A$-$D$-$E$--type potentials, namely $\C(1,1)[10]$,
$\C(1,2,4)[10]$, $\C(1,1,2,2)[5]$, $\C(1,5)[30]$, $\hbox{\,}\C(1,10,10)[30]$,
$\,\C(2,5,14)[30]$, $\C(1,5,5,7)[15]$, $\C(5,6,10)[30]$, $\C(3,5,5,5)[15]$,
and $\C(4,5,5)[20]$, while the remaining ones cannot be obtained from
tensor products of 
minimal models:
$\C(1,2)[15]$, $\C(1,4)[25]$, $\C(2,5)[35]$, $\hbox{\,}\C(4,5)[45]$,
$\,\C(2,7,19)[40]$, $\C(2,16,17)[50]$, $\C(3,19,20)[60]$,
$\C(7,10,25)[60]$, $\C(10,16,37)[90]$, and $\C(3,4,5,6)[15]$.  All but the
first three of these configurations lead to undesirable states in the
untwisted sector or to asymmetric chiral states in the twisted sectors.

The
first three models are orbifolds of one another, so it is sufficient to
consider only one of them.  The starting point that is closest to giving
the correct \pp\ is $\C(1,1,2,2)[5]$.  Here we have calculated all
orbifolds with respect to a subgroup of the nonabelian group generated by
the $\ZZ_5$ with determinant 1, the two $\ZZ_2$'s which flip the signs of
the fields with charge $2/5$, and the permutation symmetry of $X_1^5+X_1
X_3^2$ and $X_2^5+X_2X_4^2$. We find that no choice of
torsion between the generators of order two leads to the correct \pp.
In particular, it is not possible to generate two states of charge $3/5$ in
the twisted sectors.
Thus it also appears very unlikely that this invariant can be represented
as a \lgo.  We have, however, not checked whether there is a point in the
moduli space of $\C(1,1)[10]$ or of $\C(1,1,2,2)[5]$ which has a
nonabelian linear symmetry that is not a combination of phase symmetries
and permutations.

Instead of considering the chiral ring, we could also use the above
orbifolds in combination with additional minimal models to construct
string vacua and calculate the numbers \ns\ and \nsb\
of $E_6$ representations.
The corresponding numbers for the invariants (\ref y) and (\ref8) have
been calculated in \cite{sy} and \cite f, respectively. In the case of
the modular invariant (\ref y), we have checked that none of
the orbifolds is able
to reproduce these numbers in all cases, which is another proof that this
invariant 
cannot be described by an orbifold of a non-degenerate \Lg\ model with
respect to a manifest symmetry of its potential.

    \section{Automorphism invariants and the associated \lgo s}
    \subsection{Infinite series of automorphism invariants}

We now turn our attention to a class of modular invariants of
$(A_1^{(1)})^{\oplus N}$ for which we are able to identify an associated
\lgo.  The invariants to be described in this section are all
automorphism invariants, which means that each of them is due to some
fusion rule automorphism of the diagonal invariant.  Such an invariant
associates to any $N$-tuple $\vec l\equiv(l_1,l_2,...\,,l_N)$ of $A_1$
quantum numbers $l_i$ $(0\leq l_i\leq k_i)$ precisely one $N$-tuple
$\vec{\bar l}\equiv(\bar l_1,\bar l_2,...\,,\bar l_N)$ such that the map
  \beq  \rho: \ l_i \mapsto \bar l_i=\rho(l_i) \qquad \forien   \eeq
is an automorphism of the fusion rules, i.e.\ such that the fusion rule
coefficients ${\cal{N}}_{\vec l,\vec l',\vec l''}$ satisfy
  \beq  {\cal{N}}_{\!\rho(\vec l),\rho(\vec l'),\rho(\vec l'')}=
  {\cal{N}}_{\vec l,\vec l',\vec l''}\,.  \label b  \eeq
Since the invariant tensor describing the diagonal invariant is the unit
matrix
$\d_{\vec l,\vec{\bar l}}$\,, the tensor for the automorphism invariant
reads
  \beq  (Z^{(\vec k)})^{}_{\vec l,\vec{\bar l}}
  =\d_{\vec{\bar l},\r(\vec l\,)} .    \eql D

We will mainly analyse certain infinite series of invariants that are
present for any $N$ provided that $k_i\in2\ZZ$ for $i=1,2,...\,,N$.  To
describe the relevant automorphism $\rho$ explicitly, let us consider
first the case $k_i\in4\ZZ$ for $i=1,2,...\,,N$.
Denote by $I_0$ and $I_1$ the index sets of the even and odd $l_i$
values, respectively, i.e. $l_i\in2\ZZ+s$ for $i\in I_s$, $s=0,1$,
satisfying $I_0\cup I_1=\{1,2,...\,,N\}$ and $I_0\cap I_1=\emptyset$,
and set $N_s:=|I_s|$. Also allow for $s>1$ by identifying $I_s\equiv
I_{s\,\mod\,2}$. Finally set
  \beq  \sigma_i(l_i):=k_i-l_i.  \eeq
With these conventions, the automorphism $\r$ reads
  \beq  \rho(l_i)=\left\{ \begin{array}{lll}
  l_i & {\mbox{for}} & i\in I_{N_1},\\[1.5 mm]
  \sigma_i(l_i)& {\mbox{for}} & i\in I_{N_1+1}.  \end{array} \right.
  \label a  \eeq
That this indeed defines an automorphism of the fusion rules may be seen
as follows: the fusion rule coefficients are given by
${\cal{N}}_{\vec l,\vec l',\vec l''}=\prod_{i=1}^N {\cal{N}}_{l^{}_i,l_i'
,l_i''}$, where ${\cal{N}}_{l^{}_i,l_i',l_i''}$ are
the fusion rules of the
$i^{th}$ minimal model. The latter satisfy the selection rule
   \beq  {\cal{N}}_{l,l',l''}=0  \quad{\mbox{for}}\ l+l'+l''\in2\ZZ+1,
   \eeq
from which it follows that ${\cal{N}}_{\vec l,\vec l',\vec l''}$
and ${\cal{N}}_{\rho(\vec l),\rho(\vec l'),\rho(\vec l'')}$ both
vanish, except for
  \beq  {\cal{N}}_{\rho(\vec l),\rho(\vec l'),\rho(\vec l'')}=\prod_{i=1}
  ^N {\cal{N}}_{\sigma_i^{r_i}(l_i),\sigma_i^{s_i}(l_i'),\sigma_i^{t_i}
  (l_i'')}  \eeq
with
  \beq  r_i+s_i+t_i=\,0\ {\mbox{or}}\ 2 \quad \forien.  \eeq
The automorphism property (\ref b) then follows from the identity
  \beq  {\cal{N}}_{k-l,k-l',l''}={\cal{N}}_{l,l',l''}\,,  \eeq
which is obeyed by the fusion rules of the minimal models.

The fact that the map $(l_i)\mapsto(\bar l_i)$ is a fusion rule
automorphism ensures invariance under the modular transformation
$S: \t\mapsto-1/\t$; thus in order to check invariance under the
full modular group, one merely needs to verify invariance under
$T: \t\mapsto\t+1$, which on primary fields acts diagonally by the
phase $\exp(2\p\ii(\Delta-\bar\Delta))$.
In the situation at hand, $T$ invariance
is an immediate consequence of the property $\Delta_{k-l}-\Delta_l=k/4
-l/2$ of conformal dimensions of $A_1^{(1)}$ primaries. Namely, this
property implies that $\Delta_{\rho(\vec l)}-\Delta_{\vec
l}\equiv\sum_{i=1}^N
(\Delta_{\rho(l_i)}-\Delta_{l_i})$ is an integer iff the number of odd
$l_i$ that satisfy $\rho(l_i)=\sigma_i(l_i)$ is even for any tuple
$\vec l$, as is indeed fulfilled for the map (\ref a).

Let us remark that only for $N=2$ is the invariant
defined by (\ref a) `fundamental' in the sense
\cite{sy0} that it cannot be obtained by forming sums and/or products of
the invariant tensors corresponding to simpler invariants; in that case
this invariant was found in \cite{s,f}. Also
note the structural difference between even and odd values of
$N$ which shows up when
describing the left--right symmetric primary fields: for these,
either $I_1=\emptyset$, or else $N_0\in2\ZZ+N+1$ and
$l_i=k_i/2$ for all $i\in I_0$. Thus in particular for odd $N$ the
tuples $\vec l$ which give $I_0=\emptyset$ provide left--right symmetric
fields, whereas for even $N$ at least one of the $l_i$ corresponding
to a left--right symmetric field must be even.

The automorphism $\r$ can be described in a similar manner
if some of the levels $k_i$ obey $k_i\in4\ZZ+2$.
For brevity we describe only the case where $k_i\in4\ZZ$ for $i=1,2,...\,
,N-1$ and $k_N\in4\ZZ+2.$ In this situation the automorphism reads
  \beq  \begin{array}{l} \rho(l_i)\,=\,\left\{ \begin{array}{lll} l_i
  &{\mbox{for}} & i=1,2,...\,,N-1\ {\mbox{and}}\ l_i+l_N\in2\ZZ, \\[1 mm]
  \sigma_i(l_i) \ \ \;{}  & {\mbox{for}} & i=1,2,...\,,N-1\ {\mbox{and}}\
  l_i+l_N\in2\ZZ+1, \end{array}\right. \\[-2 mm] {}\\
  \rho(l_N)= \left\{
  \begin{array}{lll} l_N & {\mbox{for}} & N_1+l_N\in2\ZZ+1, \\[1 mm]
  \sigma_N(l_N)& {\mbox{for}} & N_1+l_N\in2\ZZ.  \end{array}\right.
  \end{array}  \label c  \eeq
That this is a fusion--rule automorphism follows by the same reasoning
as above. $T$ invariance now requires
that the number of those $l_i$, $i=1,2,...
\,,N-1$, for which $\rho(l_i)=\sigma_i(l_i)$, be even if $l_N$ is odd,
and odd if $l_N$ is even, and this condition is met by (\ref c).
Also note that although the description (\ref c) of the invariant
at first sight looks rather different
from the corresponding formula (\ref a) above,
it leads to the same structure
of the set of left--right symmetric primaries.

    \subsection{\Lg\ interpretation}

We want to identify an infinite series of \Lg\ theories. This implies
that we must start from the Fermat--type potential
 \beq W=\sum_{i=1}^N X_i^{k_i+2},\label{fe} \eeq
as this is the only infinite series of \Lg\ potentials
matching the central charges
(apart from potentials employing the respective $D$ invariants, but these
are related to (\ref{fe}) by an orbifold construction and thus cannot
give anything new).  It is then not useful either
to consider the \pp, as we can keep track of the charges of chiral fields as
a function of the levels and thus should identify these fields individually.
In order to see the
structure of the chiral ring corresponding to the invariant
(\ref a), observe that $\r(l_i)=\s(l_i)$ iff the number of odd $l_j$ with
$j\neq i$ is odd.  Thus fields with all $l_i$ even are in the chiral
ring.  If at least one $l_i$ is odd, then left--right symmetry implies
that there is an odd number of odd $l_j$, and that all even $l_j$ must be
equal to $k_j/2$.

Thus from the chiral ring of the potential (\ref{fe}) we need to project
out all fields $\prod_iX_{}^{l_i}$ with an even number of odd $l_i$.  The
only symmetries with $\det=\pm1$ that we have at our disposal for the
whole series of potentials are the $\ZZ_2$'s that invert the sign of a
number of fields. The subgroup with $\det=1$ 
may be described as being
generated by symmetries of the form
  \beq  \ZZ_2^{(i)}=\ZZ_2(0,\ldots,1,1,0,\ldots,0), \eql{zi}
with the 1's at the $i^{th}$ and $(i+1)^{st}$ position.  For $N$ odd this is
exactly the symmetry group we need for the projection in the untwisted
sector, as states with all numbers $l_i$ odd should survive.  The twisted
sectors come from transformations flipping the signs of an even number of
fields.  According to (\ref T) this implies a contribution $1/2-Q_i$ to
the charge of the twisted vacuum for each twisted field $X_i$.  On the
other hand, setting all torsions $\e(g,h)=1$ in (\ref H), the projection
onto invariant states keeps the odd powers of the untwisted fields.  As
$1/2-Q_i$ is exactly the charge of a chiral field with $\r(l_i)=\s(l_i)$,
we get complete agreement for the charge degeneracies in the chiral rings
of the automorphism invariant (\ref a) and the $(\ZZ_2)^{N-1}$ orbifold
of (\ref{fe}) with respect to the group generated by (\ref{zi}).

For even $N$ the states with all $l_i$ odd should be projected out.
Accordingly, we need to supplement the projection by transformations with
determinant $-1$ on the fields $X_i$, $i\le N$.  A negative determinant,
however, can always be avoided by letting the respective transformations
act on an additional trivial field $X_{N+1}$, contributing a term
$X_{N+1}^2$ to the potential.  In this way we can reduce the case of $N$
even to the previous case with $k_{N+1}=0$.  The fields with all $l_i$
odd no longer contribute to the chiral ring, as $X_{N+1}$ is a
descendant field.  As above, one can check that the complete chiral ring
works out correctly for this $(\ZZ_2)^{N}$\, orbifold.
Note that in these considerations we only need to require that $k_i\in
2\ZZ$ for $i=1,2,...\,,N$, but not necessarily $k_i\in4\ZZ$;
correspondingly, the \Lg\ interpretation found above not only applies
to invariants of the type (\ref a), but to those of type (\ref c)
etc.\ as well.

Finally, let us mention that special care is necessary for identifying
orbifolds with even order in the above models.  We illustrate this for
the case $N=2$ with the non-diagonal invariant (\ref a) of
$(A_1^{(1)})_{k_1}\oplus (A_1^{(1)})_{k_2}$, which is
described by a $(\ZZ_2)^2$\, orbifold of $X_1^{k_1+2}+X_2^{k_2+2}+X_3^2$.
Owing to the projection, the full original $\ZZ_{k_1}\ex\ZZ_{k_2}$ symmetry
is no longer present in the \LG\ representation.  Instead, however, we
have gained a new $(\ZZ_2)^2$\, symmetry that
acts on the twisted sectors. Now consider, for example, the
$\ZZ_2$ generator $g_1$ of $(A_1^{(1)})_{k_1}$, which acts non-trivially on
the fields with $l_1$ odd and $l_2=k_2/2$ in the chiral ring. These fields
correspond to the sector twisted by the transformation $t_2$, which acts
non-trivially on $X_2$ and $X_3$. If we now want to orbifoldize a model
with this invariant by a symmetry $\s_1$ that acts like $g_1$ in the
first factor of the tensor product, then we need to make $\s_1$ act
non-trivially on the twisted vacuum $|t_2\>$.  This can be achieved by
introducing a discrete torsion $\e(\s_1,t_2)=-1$ (compare formula (\ref
H)).  Another possibility, which keeps the way open for a geometric
interpretation of the string vacuum, would be to simulate the effect of
this torsion by introducing two additional trivial fields and letting
$\s_1$ act on one and $t_2$ act on both of them (such that all
determinants
remain positive).\footnote{~In fact, within the \LG\ framework, any
 $\ZZ_2$ torsion between two generators can be simulated by introducing
 three additional trivial fields, with the two generators 
 acting on different pairs of these fields.}

     \subsection{An exceptional automorphism invariant}

As a last example we consider an exceptional invariant that is again of the
automorphism type, but now the automorphism is not with respect to
$(A_1^{(1)})^{\oplus N}$ fusion rules, but rather with respect to the
fusion rules of an extended chiral algebra. The invariant arises for
$N=2$ and $k_1=3$, $k_2=28$; it reads \cite f
  \beq  Z_{}^{(3,28)} = | 0,\!\tilde0 |^2_{} \oplus | 1,\!\tilde6 |^2_{}
  \oplus | 2,\!\tilde6 |^2_{} \oplus | 3,\!\tilde0 |^2_{}
  \oplus \left[ (0,\!\tilde6)\otime(2,\!\tilde0)^*_{}
  \oplus (1,\!\tilde0)\otime(3,\!\tilde6)^*_{}
  \oplus \mbox{c.c.} \right] ,    \label 3 \eeq
where
  \beq  \tilde0\equiv0\oplus10\oplus18\oplus28, \qquad
  \tilde6\equiv6\oplus12\oplus16\oplus22. \eeq
The extension of the chiral algebra corresponds to the conformal
embedding of $(A_1^{(1)})^{}_{28}$ in $(G_2^{(1)})^{}_1$, as in the
case of the exceptional $E_8$ type $A_1^{(1)}$\, invariant.

In order to find a \lgo\ that corresponds to the invariant (\ref3),
we start by writing down the corresponding \pp.
It reads
\beq P^{(3,28)}(t)=(1+t^{1/3})(1+t^{2/5}+4t^{3/5}+t^{4/5}+t^{6/5}). \eql{P3}
Let us express this \pp\ in terms of the variables
$x=t^{1/5}$, $y=u^6=t^{6/30}$, and $z=u^{10}=t^{10/30}$; this is
suggested by the \Lg\ potential $Y^5+Z^3$
for the exceptional invariant of the level--28 theory,
whose building blocks are used in $Z^{(3,28)}$.
In these variables the \pp\ becomes
  \beq  \begin{array}{l} P^{(3,28)}(t)\,= (1+x^3)(1+u^{10}+u^{18}+u^{28})
  +(x+x^2) (u^6+u^{12}+u^{16}+u^{22}) \\[1.6 mm] \qquad\qquad\;
  =(1+z) [(1+x^3)(1+y^3)+(x+x^2)(y+y^2)].  \end{array} \eeq
We thus start from the potential \beq X^5+Y^5+Z^3 \eeq and 
try to use a twist in the (X,Y) sector to get the second factor in the
expression (\ref{P3}) for the \pp.  To avoid a contribution $t^{1/5}$ to the
polynomial, we need to eliminate the fields $X$ and $Y$ and any
linear combination thereof from the untwisted sector. As the determinant
of the twist should be $\pm1$, we have to use the $\ZZ_5$ which acts as
$(X,Y,Z)\mapsto(\l X,\l^4Y,Z)$ with $\l^5=1$.  From the untwisted sector
we now get exactly the even powers of $t^{1/5}$ on the right--hand side of
(\ref{P3}), whereas the four twisted sectors contribute $t^{3/5}$ each.
Thus the $\ZZ_5{(1,4,0)}$\, orbifold of $X^5+Y^5+Z^3$ reproduces the
correct \pp\ for the automorphism invariant (\ref 3).

\begin{table}   \centering \caption{\em Spectra of models based on 
exceptional \mbox{$(A_1^{(1)})^{\oplus N}$} invariants}
 \vspace{7 mm}
\begin{tabular}{|rr|r|r|r|r|r|}   \hline &&& &&& \\[-3.6 mm]
\multicolumn{1}{|c}{\mbox{$\#$}} & \multicolumn{1}{c|}{\sf invariant}
& \multicolumn{1}{c|}{\mbox{$n^{}_{27}$}}
& \multicolumn{1}{c|}{\mbox{$n^{}_{\overline{27}}$}}
& \multicolumn{1}{c|}{\mbox{$n^{}_1$}}
& \multicolumn{1}{c|}{\mbox{$n^{}_g$}} & \multicolumn{1}{c|}{$\chi$}
\\[-3.7 mm] &&& &&&\\ \hline\hline &&& &&& \\[-3.4 mm]

  50 & 1 2 4 4-10     &  43 &   7 & 232 &  4 &  $-$72     \\[.7 mm]
 127 & 2 4 12-82      &  51 &  57 & 447 &  3 &     12     \\[.7 mm]
 130 & 2 4 16-34      &  66 &  24 & 353 &  3 &  $-$84     \\[.7 mm]
 131 & 2 4 18-28      &  36 &  42 & 327 &  3 &     12     \\[.7 mm]
 133 & 2 4-22 22      &  74 &  14 & 341 &  3 & $-$120     \\[.7 mm]
 136 & 2 5 12-26      &  59 &  23 & 327 &  3 &  $-$72     \\[.7 mm]
 138 & 2 6  8-38      &  38 &  38 & 327 &  3 &      0     \\[.7 mm]
 138 & 2  6-8 38      &  38 &  38 & 327 &  3 &      0     \\[.7 mm]
 143 & 2 8  8-18      &  66 &  12 & 315 &  3 & $-$108     \\[.7 mm]
 144 & 2 8 10-13      &  18 &  42 & 247 &  3 &     48     \\[.7 mm]
 160 & 4 6  4-22      &  41 &  17 & 243 &  3 &  $-$48     \\[.7 mm]
 163 & 4 4-10 10      &  62 &   8 & 263 &  3 & $-$108
\\[-3.9 mm] &&& &&& \\ \hline\hline &&& &&& \\[-3.4 mm]
  17 & \ 1 1 1 4-4-4  &  51 &   3 & 213 &  5 &  $-$96     \\[.7 mm]
  54 & 1 4 4-4-4      &  37 &   7 & 200 &  4 &  $-$60     \\[.7 mm]
  59 & 2 2 4-4-4      &  12 &  30 & 215 &  5 &     36     \\[.7 mm]
  95 & 1 8-16-88      &  55 &  61 & 447 &  3 &     12     \\[.7 mm]
  98 & 1 8-28-28      &  95 &  17 & 419 &  3 & $-$156     \\[.7 mm]
 111 & 1 12-12-40     &  60 &  30 & 345 &  3 &  $-$60     \\[.7 mm]
 117 & 1 16-16-16     & 101 &  11 & 401 &  3 & $-$180     \\[.7 mm]
 153 & 3 4-8-28       &  39 &  27 & 277 &  3 &  $-$24     \\[.7 mm]
 158 & 3 8-8-8        &  67 &   7 & 267 &  3 & $-$120     \\[.7 mm]
 159 & 5 4-4-40       &  38 &  32 & 299 &  3 &  $-$12     \\[.7 mm]
 161 & 7 4-4-16       &  50 &  14 & 261 &  3 &  $-$72     \\[.7 mm]
 162 & 13 4-4-8       &  29 &  23 & 223 &  3 &  $-$12
\\[-3.9 mm] &&& &&& \\ \hline\hline &&& &&& \\[-3.4 mm]
  54 & 1 4-4-4-4      &  37 &   7 & 200 &  4 &  $-$60
\\[-3.9 mm] &&& &&& \\    \hline\end{tabular}
 \end{table}

   \subsection{Some string spectra}

We have verified that the Gepner--type models that
employ the invariants described in subsections 4.1 and 4.3 and the string
compactifications obtained from the corresponding \lgo s possess
identical spectra of massless ($E_6$ non-singlet) fields, thus further
confirming the identification of the respective theories.  In the case of
the invariant (\ref3), as well as for (\ref D) with $N=2$ and
$k_1,\,k_2\in4\ZZ$, these spectra have already been listed in \cite{f}.
The spectra for a few other models employing invariants of the type (\ref
D) are listed in table 1, namely for some models with $N=2$ and
$k_1\in4\zz,\;k_2\in4\ZZ+2$, for all possible $c=9$ models with $N=3$ and
$k_1,k_2,k_3\in4\ZZ$, and for the single $c=9$ model with $N=4$ and
$k_1,k_2,k_3,k_4\in4\ZZ$.

The notation in table 1 is as follows. The first column contains the
number which in \cite{num} has been associated to
the relevant combination $(k_1,k_2,\ldots)$ of levels of the affine
algebras. In the second column we give the chosen invariant, with $k$
standing for the $A$ type invariant at level $k$, and with $k_1$-$k_2$
denoting the invariant (\ref D) with $N=2$ and levels $k_1,\,k_2$, etc.
The next four columns contain the spectrum, i.e.\ the
numbers \ns, \nsb\ and $n_1$ of massless matter fields in the $27$,
$\overline{27}$ and singlet representations of $E_6$, respectively, as well as
the number $n_g$ of gauge bosons that are present in addition to those of
$E_6$. The number in the last
column is the Euler number $\chi=2(\nsb-\ns)$.


\begin{table}  \centering \caption{\em Spectra of orbifolds of
the model {\rm \# 127} with invariant {\rm2\,4-12\,82}} \vspace{7 mm}
\begin{tabular}{|l|r|r|r|r|r|}   \hline && &&& \\[-3.6 mm]
\multicolumn{1}{|c|}{\sf twist}
& \multicolumn{1}{c|}{\mbox{$n^{}_{27}$}}
& \multicolumn{1}{c|}{\mbox{$n^{}_{\overline{27}}$}}
& \multicolumn{1}{c|}{\mbox{$n^{}_1$}}
& \multicolumn{1}{c|}{\mbox{$n^{}_g$}} & \multicolumn{1}{c|}{$\chi$}
\\[-3.7 mm] && &&&\\ \hline\hline && &&& \\[-3.4 mm]

{\bf1}               &  42 &  48 & 357 &  3 &     12     \\[.7 mm]
\zz2   (0,1,1,0)     &  49 &  49 & 375 &  3 &      0     \\[.7 mm]
\zz2   (1,0,1,0)     &  33 &  39 & 325 &  3 &     12     \\[.7 mm]
\zz2   (1,1,0,0)     &  39 &  33 & 325 &  3 &  $-$12     \\[.7 mm]
\zz4   (1,0,0,3)     &  30 &  96 & 479 &  3 &    132     \\[.7 mm]
\zz4   (1,0,2,1)     &  76 &  40 & 440 &  4 &  $-$72     \\[.7 mm]
\zz4   (1,2,0,1)     &  58 &  34 & 371 &  3 &  $-$48   
\\[-3.9 mm] && &&& \\     \hline\end{tabular}
 \end{table}

As already mentioned,
we checked that the \Lg\ results agree with all spectra given in \cite{f}
whenever we have a \Lg\ interpretation of the respective invariant.
We have done so not only for the models themselves, but for further
orbifoldized versions of them as well. Note that
this requires the use of appropriate discrete torsions, as explained at
the end of section 4.2.
Actually some of the orbifold results
have been reproduced incorrectly in table~3 of \cite f.
Therefore we also list, in table\ 2, the correct
spectra for these theories. The model in question is the one numbered as
\# 127, with invariant
2\,4-12\,82. 
The
notation used for the modded out symmetry is as in formula (\ref{co}).

\section{Outlook}

In this paper we have described a recipe of how to search
systematically for the \Lg\ interpretation of any given modular
invariant for tensor products of minimal $n=2$ superconformal models.
We have applied this procedure in a case-by-case analysis
to various invariants for which an
associated \lgo\ could be identified. On the other hand,
we were also able to use
the method to prove that the particular invariant (\ref y) cannot be
described in terms of a \lgo\ (with respect to a manifest linear
symmetry).

Of course it would be desirable to understand at a more
fundamental level why in some cases such a correspondence exists while
in other cases it does not.  In this context we note that all invariants
described in section 4 are of the automorphism type; thus
maybe at least all invariants of this particular type possess a \Lg\
interpretation.  There is, however, no obvious connection between the
$\ZZ_2$ group corresponding to the fusion rule automorphism and the
$\ZZ_2$ symmetries (\ref{zi}) modded out in the orbifold construction.

Let us also mention that in the special situations we are considering the
possible dependence on the moduli of potentials with $c\ge3$ does not
pose any problem for a unique identification of a \Lg\ theory, as these
moduli are fixed by discrete symmetries.

In a sense, the automorphism invariant (\ref a) is a generalization of
the $D$ invariant of minimal models.  It is thus tempting to look for a
non-linear transformation like the one that relates the $\ZZ_2$ orbifold
representation of that invariant to an (untwisted) \Lg\ model.  For $N=2$
and $W=X^{2a}+Y^{2b}+Z^2$, for example, $X'=X^2$, $Y'=Y^2$ and
$Z'=Z/(XY)$ indeed provides a transformation with the required
$(\ZZ_2)^2$ identification and with constant determinant.  The resulting
potential $(X')^a+(Y')^b+X'Y'(Z')^2$, however, is degenerate.  Although
in some cases the configuration determined by this
polynomial is non-degenerate, the spectra calculated from a regularizing
deformation of such polynomials
turn out wrong.

Our findings corroborate with recent indications that the space of \lgo s
is not mirror--symmetric \cite{algo}
to diminish 
the hope 
that \Lg\ models are of much use for a classification of
$n=2$ superconformal field theories.
Still, it appears worth while to apply our ideas to other constructions
such as $n=2$ coset models \cite{ks}.
In any case, a \Lg\ representation, if it exists, provides an extremely
efficient computational framework. In addition, one may get an idea of
the nature of the limitations of \Lg\ models. 

}

\end{document}